\documentclass[12pt,a4paper,nofootinbib,superscriptaddress]{revtex4-1}

\usepackage{ulem}
\usepackage{color}
\usepackage{graphicx}
\usepackage{hyperref}   
\usepackage{appendix}
\usepackage{epsfig}
\usepackage{epstopdf}
\usepackage{amsmath}
\usepackage{subfig}
\usepackage{comment}
\usepackage[dvipsnames]{xcolor}
\usepackage{epsfig}
\usepackage{appendix}
\usepackage{graphics}
\usepackage{graphicx}
\usepackage{color}
\newcommand{\jpsi}{J / \psi}

\newcommand{\old}[1]{}

\newcommand{\be}{\begin{equation}}
\newcommand{\ee}{\end{equation}}
\newcommand{\ba}{\begin{eqnarray}}
\newcommand{\ea}{\end{eqnarray}}
\newcommand{\bi}{\begin{itemize}}
\newcommand{\ei}{\end{itemize}}

\newcommand{\nn}{\nonumber}

\begin{document}
\title{Heavy quark potential and LQCD based quark condensate at finite magnetic field}
\author{Indrani Nilima}
\affiliation{Indian Institute of Technology Bhilai, GEC Campus, Sejbahar, Raipur - 492015, Chhattisgarh, India}
\author{Aritra Bandyopadhyay}
\affiliation{Guangdong Provincial Key Laboratory of Nuclear Science, Institute of Quantum Matter, South China Normal University, Guangzhou 510006, China}
 \affiliation{Guangdong-Hong Kong Joint Laboratory of Quantum Matter, Southern Nuclear Science Computing Center, South China Normal University, Guangzhou 510006, China}
 \author{Ritesh Ghosh}
\email{ritesh.ghosh@saha.ac.in}
\affiliation{
	Theory Division, Saha Institute of Nuclear Physics, HBNI, \\
	1/AF, Bidhannagar, Kolkata 700064, India}
\affiliation{
	Homi Bhabha National Institute, Anushaktinagar, \\
	Mumbai, Maharashtra 400094, India}
 \author{Sabyasachi Ghosh}
\affiliation{Indian Institute of Technology Bhilai, GEC Campus, Sejbahar, Raipur - 492015, Chhattisgarh, India}
%

%\date{}
%\maketitle
\begin{abstract}
   
   In the present work, we have studied heavy quarkonia potential in hot and magnetized quark gluon plasma. Inverse magnetic catalysis (IMC) effect is incorporated within the system through the magnetic field modified Debye mass by modifying the effective quark masses. We have obtained the real and imaginary part of the heavy quark potential in this new scenario. After the evaluation of the binding energy and the decay width we comment about the dissociation temperatures of the heavy quarkonia in presence of magnetic field.
\end{abstract}
\maketitle
\section{Introduction}
A lot of information is being provided by the ongoing relativistic heavy-ion collisions (HIC) in respect of deconfined state of matter, i.e. Quark Gluon Plasma (QGP). 
%Currently ongoing experimental programs in RHIC/LHC has the aim to study the properties of QGP at high temperatures. 
QGP at sufficiently high temperature behaves like a weakly interacting gas of quarks and gluons, which can also be studied perturbatively by using hard thermal loop (HTL) resummation~\cite{Weldon:1982aq, Braaten:1989mz,Frenkel:1989br,Braaten:1991gm} techniques, apart from the first principle lattice QCD estimations. Depending on the non-centrality, HIC can also produce a very strong magnetic field in the direction perpendicular to the reaction plane~\cite{Kharzeev:2007jp,Kharzeev:2007jp,Skokov:2009qp}. At the RHIC energies, the estimated strength of the magnetic field is around $B \sim m_{\pi}^2 \equiv 10^{18}$ Gauss whereas at the LHC the estimated strength is around $B \sim 15m_{\pi}^2 \equiv 1.5 \times 10^{19}$ Gauss \cite{Kharzeev:2007jp,Skokov:2009qp}, where $ m_{\pi} $ is the pion mass. The issue - whether this large initial magnetic field will decay very fast~\cite{McLerran:2013hla} or slow~\cite{Tuchin:2010gx,Das:2017qfi} is still probably an open problem. Studies have also shown that such an extensive magnetic field might even have survived from the very early stages of universe~\cite{Vachaspati:1991nm,Grasso:2000wj}. These possibilities provide the opportunity to revisit the entire QGP phenomenology in presence of magnetic field, and in the present work we have attempted the same with heavy quark phenomenology. 

%In the quarks / antiquarks, the magnetic field effects enter in the degrees of freedom through various Landau levels, which signifies the quantization of the transverse momenta in the dispersion relation. Inclusion of the external magnetic field in the theoretical studies of quark matter have produced a large body of work and various novel and interesting features in recent years. One of the most interesting aspects of a magnetized medium is the discovery of a pair of contrasting effects, i.e. 
According to recent lattice quantum chromodynamic (LQCD) calculation~\cite{Bali1,Bali2}, non-zero QCD vaccum at finite temperature and magnetic field can face both
magnetic catalysis (MC) and inverse magnetic catalysis (IMC). MC shows the low temperature enhancements in the values of quark condensates with increasing magnetic field and extensively studied through lattice QCD and effective models. On the other hand, IMC shows a decreasing behaviour of the condensates with increasing magnetic field close to the transition temperature. IMC was first discovered by lattice QCD simulations using physical values of pion mass and for the light quarks. Since then there have been many efforts to implement IMC in the effective models. In our present work, we have implemented this important IMC effect in our calculation through the constituent quark mass generated by the lattice QCD simulations, which incorporates the complex and non-monotonic temperature and magnetic field dependence within the quark condensates.

One of the useful probe of the QGP formation is heavy quarkonium which is a bound state of $ Q\bar{Q} $ pair~\cite{McLerran:1986zb, Back:2004je}. After the discovery of $J/\psi$ ( a bound state of $c {\bar c}$), ~\cite{Aubert:1974js, Augustin:1974xw}, in 1974, a large number of excellent articles have been published that proposed several essential refinements in the study of heavy quark potential. The first work to study the quarkonia  at finite temperatures using potential models have been done by Karsch, Mehr, and Satz~\cite{mehr}. Subsequently another pioneering work to study the dissociation of quarkonia due to the color screening in the deconfined medium with finite temperature, was carried out by Matsui and Satz~\cite{Matsui:1986dk}. In recent years various studies have been executed to see the impact of magnetic field on heavy quark phenomenology~\cite{Alford:2013jva,Marasinghe:2011bt,Cho:2014exa,Guo:2015nsa,Bonati:2015dka,Rougemont:2014efa,Dudal:2014jfa,Sadofyev:2015hxa,Machado:2013rta,Machado:2013yaa,Gubler:2015qok,Fukushima:2015wck,Das:2016cwd,Singh:2017nfa,Kurain:2019,Hasan:2017fmf,Bandyopadhyay:2021zlm}.
Specifically speaking, the effects of the external magnetic field on the quarkonia production have been discussed in Refs.~\cite{Machado:2013rta,Guo:2015nsa}. In Ref.~\cite{Das:2016cwd}, the authors have studied the directed flow of charm quarks which is considered as an efficient probe to characterize the evolving magnetic field produced in ultra-relativistic HIC. In Ref.~\cite{Machado:2013yaa} the authors have investigated QCD sum rules in calculation of the mass of heavy mesons to estimate the modification of the charged B meson mass, ($ m_B $), in the presence of an external Abelian magnetic field. The momentum diffusion coefficients of heavy quarks, in a strong magnetic field within the lowest Landau level (LLL) approximation, along the directions parallel and perpendicular to $B$, at the leading order in QCD coupling constant have been computed within and beyond the static limit of the heavy quarks, respectively in Refs.~\cite{Fukushima:2015wck} and \cite{Bandyopadhyay:2021zlm}.

The physics about the fate of quarkonia at zero temperature can be understood with the help of non-relativistic potential models. Masses of heavy quarks ($ m_Q $) are much larger than QCD scale ($ \Lambda_{QCD} $) and velocity of the quarks in the bound state is small, $v\ll 1$ \cite{Lucha:1991vn}. Hence, to understand the binding effects in quarkonia generally one uses the Cornell potential which belongs to the family of the non-relativistic potential models~\cite{Eichten:1979ms} and can be derived directly from QCD using the effective field theories (EFTs)~\cite{Eichten:1979ms,Lucha:1991vn,Brambilla:2004jw}. 
Refs.~\cite{Singh:2017nfa,Hasan:2017fmf} has studied the effect of a strong magnetic field on the heavy quark complex potential with the lowest Landau level (LLL) approximation. In this regard, present work has moved beyond LLL estimation and considered all Landau level summation, which is valid for the entire range of magnetic field from weak to strong. This is one of the new ingredients of the present work. The main goal of the present work though is to incorporate the IMC effect in the heavy quark potential through the effective quark masses. These two components are mainly introduced within the standard formalism of heavy quark potential~\cite{Agotiya:2008ie,Thakur:2012eb,Thakur:2013nia,Kakade:2015xua,Agotiya:2016bqr}, which is the sum of both Coulomb and string terms~\cite{Eichten:1974af}.

%For obtaining the real and imaginary parts of
%medium modified heavy quark potential we will first obtain the real and imaginary parts of the effective gluon propagator, which in turn gives the real and imaginary parts of the dielectric permittivity. Now the real and imaginary parts of the dielectric permittivity will give the real and imaginary parts of the complex
%heavy quark potential~\cite{Agotiya:2008ie,Thakur:2012eb,Thakur:2013nia,Kakade:2015xua,Agotiya:2016bqr}. For calculating the binding energy of heavy quarkonia the real part of the potential is used in the Schrodinger equation whereas the imaginary part is used to calculate the thermal width. Finally we will obtain the dissociation temperatures of heavy quarkonia in presence of magnetic field by using the criterion that thermal width is overcome by twice of the real part of the binding energy. We will also study how the dissociation temperatures get affected in presence of magnetic field. In the present investigation, we focus on the modification of the heavy quark potential in the presence of strong magnetic field through the modification of the Debye screening mass, which has been shown to be the most essential component of the heavy quark potential. Here our main motivation is to show the IMC effect of finite magnetic field on the Debye sceening mass induced from the real part of the potential and  Landau-damping induced from thermal width which is obtained from the imaginary part of the heavy quark potential. 

The paper is organized as follows. In section~\ref{sec2},  we will discuss the basic formalism of our present work which includes discussions about the real and imaginary parts of the heavy quark potential, decay width, binding energy and the Debye screening mass. In section~\ref{sec3} we will show our results for the same as well as find out the dissociation temperatures and discuss their magnetic field dependence. Finally in section~\ref{sec4}, we shall conclude the present work.

\section{Formalism}
\label{sec2}
In this section, we have described the entire formalism required for our current study. In the first subsection (\ref{HQP}), we have addressed the standard framework of the heavy quark potential in presence of an external magnetic field, and established the connection with the gluon propagator through the dielectric permittivity. In the process, a temperature and magnetic field dependent Debye mass enters into the heavy quark potential through the gluon propagator. This Debye mass is calculated from semi-classical transport theory, whose magneto-thermodynamical phase space can be obtained by projecting the temperature and magnetic field dependent condensates from the lattice quantum chromodynamics (LQCD) calculation, which incorporate the effects of both MC and IMC. This part is discussed in details in subsection (\ref{sec:mD}), which is the main motive our present study. Next in subsection (\ref{sec:G_E}), we have discussed about the framework of thermal width and binding energy of heavy quarkonia (e.g. $\jpsi$ or $\Upsilon$) from imaginary and real part of heavy quark potential respectively.

\subsection{In-medium heavy quark potential in presence of magnetic field}
\label{HQP}
 Let us start our discussion with the full Cornell potential~\cite{Eichten:1978tg,Eichten:1979ms}, that 
contains the Coulombic as well as the string part given as,
\ba
{\text V(r)} = -\frac{\alpha}{r}+\sigma r~.
\label{eq:cor}
\ea
Here, $r$ is the effective radius of the corresponding quarkonia state, $\alpha$ is  the strong coupling constant given by $({\alpha}={\alpha}_sC_F=\frac{g_s^2C_F}{4\pi}; C_F=4/3)$
and $\sigma$ is the string tension.

The Fourier transform of ${\text V(r)}$ is
\ba
{\bar{\text V}}(k)= -\sqrt\frac{2}{\pi}\bigg(\frac{\alpha}{k^2}+2\frac{ \sigma}{k^4}\bigg)~.
\ea
The assumption given
 in Ref.~\cite{Agotiya:2008ie} has been followed which says that, the in-medium modification can be obtained
 in the Fourier space by dividing the heavy-quark potential from the medium dielectric permittivity, $\epsilon({\bf k},T,eB)$ as,
 \ba
 {\tilde{V}}(k,T,eB)=\frac{{\bar{\text V}}(k)}{\epsilon(k,T,eB)}~,
 \ea
 where $\epsilon(k,T,eB)$ can be obtained from the static limit of the longitudinal part of gluon
self-energy~\cite{schneider}.
 So, information of temperature $T$ and magnetic field $eB$ (in M$^2$ dimension) are entered through this medium dielectric permittivity $\epsilon({\bf k},T,eB)$. 
 By making the inverse Fourier transform, we can obtain the modified potential at finite $T$ and $eB$ as,
 \ba 
\bar{{\text V}}(r,T,eB)= \int \frac{d^3\mathbf{k}}{(2\pi)^{3/2}}(e^{i\mathbf{k} \cdot \mathbf{r}}-1)\tilde{V}(k,T,eB)~.
 \label{eq:V}
 \ea
 
Next, the dielectric permittivity can be obtained in the static limit, in the Fourier space, from the temporal component 
 of the propagator ($\Delta^{00}$) as ~\cite{Weldon:1982aq,schneider},
 \ba
\epsilon^{-1}({\bf k},T,eB) = -\lim_{\omega \to 0}k^2\Delta^{00}(\omega,{\bf k},T,eB).
\label{eq:eps}
\ea
%%%%%%%%%%%%%%%%%%%%%%%%%%%%%%%%%%%%%%%%%%%%%%%%%%%%%%%%%%%%%%%%%%%%

Now, to obtain the real part of the inter-quark potential in the static limit, the temporal component 
 of  real part of the retarded propagator in the Fourier space is demanded, which is given as
  \ba
  Re[\Delta^{00}(\omega = 0,{\bf k},T,eB)] &=&\frac{-1}{k^2+m_{D}^2(T,eB)}
  \ea
 The imaginary part of the same can be derived from
the imaginary part of the temporal component of symmetric propagator in the static limit which is given as~\cite{Sym_Prop}
\ba
Im[\Delta^{00}(\omega = 0,{\bf k},T,eB)]& =& \pi~ T~ m_{D}^2\bigg\{\frac{-1}{k\{k^2+m_{D}^2(T,eB)\}^2}\nn\Big\} 
\label{eq:Im_delta}
\ea
Thus in the short-distance limit ($ \hat{r} \ll 1$),
the sum of Coulomb and string term gives the real and imaginary part of the potential in terms of modified coordinate space $\hat{r}=rm_D$ given as~\cite{Agotiya:2016bqr,V_mD1,V_mD2}
%\label{repotn}
\begin{eqnarray}
\label{pis}
Re V(\hat{r};T,B)=\left(\frac{2\sigma}{m_D}-\alpha m_D\right)\frac{e^{-\hat r}}{\hat r}
-\frac{2\sigma}{m_D \hat r}+\frac{2\sigma}{m_D}-\alpha m_D~,
\end{eqnarray}
and
\begin{eqnarray}
Im V (\hat{r};T,B)=T\left(\frac{\alpha {\hat r^2}}{3}
-\frac{\sigma {\hat r}^4}{30m_D^2}\right)\log(\frac{1}{\hat r})
\label{imis}
\end{eqnarray}
respectively. Reader can notice that the information of $T$ and $B$ are entering through Debye mass $m_D=m_D(T,B)$, whose mathematical derivation is addressed in next subsection.

\subsection{Debye mass in presence of magnetic field}
\label{sec:mD}
Debye screening mass is an important observable in the context of heavy ion collisions which also acts as a QGP signature through heavy quarkonia (i.e. $J/\Psi$ and $\Upsilon$) suppression. Debye screening mass can be directly evaluated from the temporal component of the gluon self energy tensor ($\Pi_{00}(p_0,\vec{p})$) by employing the static limit ($|\vec{p}|=0, p_0\rightarrow 0$) through a perturbative order by order evaluation. On the other hand one can also determine the Debye screening mass through the semi-classical transport theory by using the relation~\cite{Carrington,Chandra_Ravi,Agotiya:2016bqr}
\begin{equation}
    m_D^2 = g_s^2 C_{q/g} \int \frac{d^3p}{(2\pi)^3} ~\frac{d}{dE} f_{q/g}(E),
\end{equation}
where $g_s^2 \equiv 4\pi\alpha_s$ is the strong coupling constant, $C_{q/g}$ 
%(...{\bf Expressions are not clear....will discuss on it}...) 
are the Casimir constants for quarks and gluons and $f_{q/g} = \frac{1}{e^{\beta E}\pm 1}$ are their respective distribution functions - Fermi-Dirac (FD) for quark and Bose-Einstein (BE) for gluons. For an ideal non interacting QGP medium, one can readily trace back the leading order hard thermal loop (HTL) expression of the Debye mass, i.e. $m_D^2=\frac{g_s^2T^2}{3} (N_c + N_f/2)$. 
There are several investigations on the Debye screening masses of the QGP as a function of the magnetic field from the temporal component of the gluon polarization in perturbative QCD (pQCD) calculation~\cite{Bandyopadhyay:2016fyd,Bonati:2017uvz,Singh:2017nfa,Kurian:2017yxj,Mitra:2017sjo,Kurain:2019,Karmakar:2019tdp}, whose equivalence anatomy in semi-classical transport theory has been shown by Ref.~\cite{SG_VC}. The gluonic distribution function remains unchanged in presence of an external anisotropic magnetic field along the $z$ direction ($\vec{B}=B\hat{z}$), whereas the quark distribution function gets modified to: 
\begin{equation}\label{2}
f^l_{q}=\dfrac{1}{\exp{(\beta E_f^l)}+1},
\end{equation}    
where the Landau quantized dispersion relation reads as $E^l_f=\sqrt{p_{z}^{2}+m^{2}_f+2l |q_f eB|}$, with $l=0,1,2,..$ being the number of Landau levels and $q_f=+\frac{2}{3}, -\frac{1}{3}$ being the fractional charge of the $u$ and $d$ 
quarks respectively. Again, in a magnetized medium the phase space quantization~\cite{Bruckmann:2017pft,Tawfik:2015apa,
Gusynin:1995nb} can be represented as 
\begin{equation}\label{3}
\int{\dfrac{d^{3}p}{(2\pi)^{3}}}\rightarrow
\dfrac{\mid q_feB\mid}{2\pi}\sum_{l=0}^{\infty}\int{\dfrac{dp_{z}}{2\pi}}(2-\delta_{0l}).
\end{equation} 
Incorporating these, the expression for the Debye screening mass $m_D$ for a magnetized QGP medium becomes,
\begin{equation}\label{4}
m_{D}^{2}=g_s^2 T^2 \frac{N_c}{3} +\dfrac{g_s^2|q_feB|}{\pi^2T}
\int_{0}^{\infty} dp_z\sum_{l=0}^{\infty}(2-\delta_{l0})f^l_q(1-f^l_q).
\end{equation}
%%%%%%%%%%%%%%%%%%%%%%%%%%%%%%%%%%%%%%%%%%%%%%%%%%%%%%%%%%%%%%%%%%%
\begin{figure} %[tbh]
\begin{center}
\includegraphics[width=7.7cm]{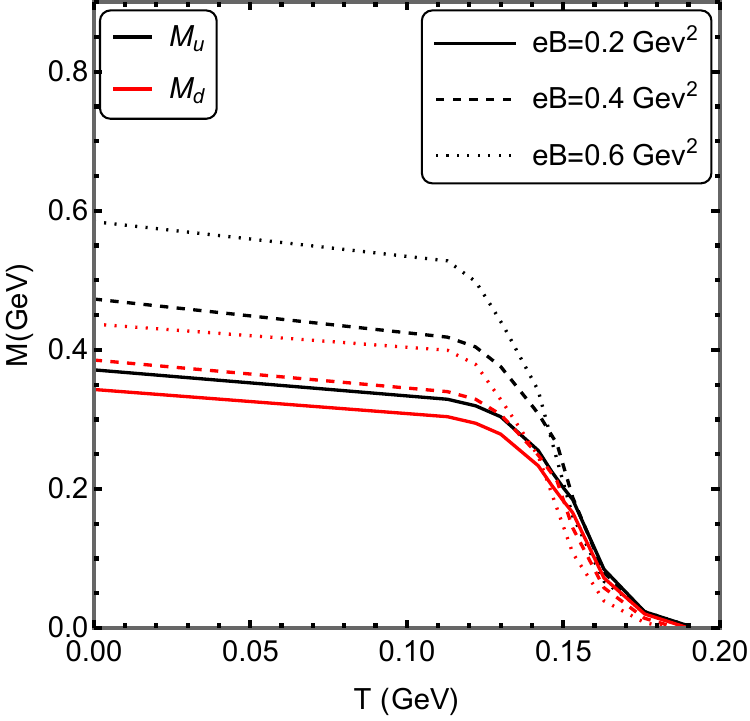} \includegraphics[width=7.7cm]{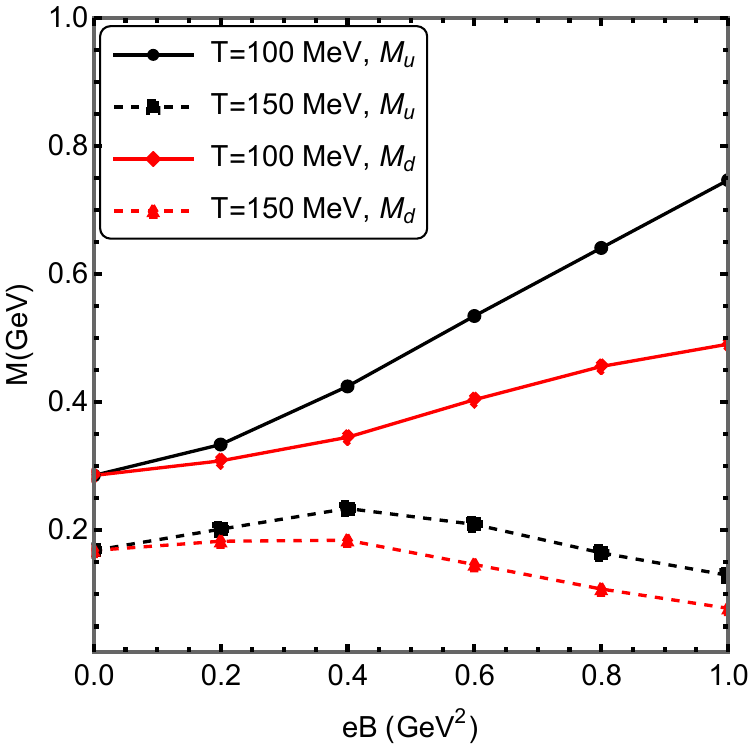}
\caption{Variation of the constituent quark masses ($M_u$ and $M_d$) with temperature for different values of magnetic field $eB=0.2$ (solid line), $0.4$ (dashed line), $0.6$ (dotted line) GeV$^2$ in left panel and with magnetic field for different values of $T=100$ (solid line), $150$ (dash line) MeV in right panel. Black and Red curves denotes respectively $u$ and $d$ quarks.}
\label{fig:M_TB}
\end{center}
\end{figure}
%%%%%%%%%%%%%%%%%%%%%%%%%%%%%%%%%%%%%%%%%%%%%%%%%%%%%%%%%%%%%%%%%%%%
Now, in the earlier studies of the Debye screening mass in a magnetized medium~\cite{Bandyopadhyay:2016fyd,Bonati:2017uvz,Singh:2017nfa,Kurian:2017yxj,Mitra:2017sjo,Kurain:2019,Karmakar:2019tdp,SG_VC}, one of the important aspect of QCD at finite magnetic field have not been considered. It is Inverse Magnetic Catalysis (IMC) phenomenon near quark-hadron phase transition temperature, recently revealed from lattice QCD (LQCD) calculations~\cite{Bali1,Bali2}. Their observed phenomenon and connected physics can be described as follows. We know that QCD vacuum or quark condensate in vacuum is non-zero, for which current quark mass $m_f\sim 5-10$ MeV get an effective constituent quark mass $M_f$ (more than $300$ MeV)  and fusion of three constituent quarks can able justify the origin of nucleon mass ($\sim 939$ MeV). Now, in presence of magnetic field, condensate value in vacuum or $T=0$ enhances, which is known as magnetic catalysis (MC) phenomenon. LQCD calculations~\cite{Bali1,Bali2} found this MC effect within low $T$ domain but near transition $T$ range, its inverse nature is observed, which is called as IMC. To capture this detailed MC and IMC aspects of QCD through $T$ and $B$ dependent quark condensate and constituent quark mass $M_f(T, B)$ in the present work, we have redefined the dispersion relation in a magnetized medium as 
\begin{equation}
    \bar{E}_f^l = \sqrt{p_z^2+2l|q_f eB|+M_f(T,eB)^2},
\end{equation}
where $M_f(T,eB)$ represents the effective constituent mass in terms of LQCD~\cite{Bali1,Bali2} predicted normalized quark condensate $\langle q{\bar q}\rangle_f(T,eB)$, which varies from 1 to 0 during the transition from hadronic to quark temperature regions for $eB=0$ case. Corresponding constituent quark mass will vary from $M_f(T=0,eB=0)$ to $m_f(T,eB=0)$ and owing to effective QCD model, we can build a connecting relation between $M_f(T,eB)$ and LQCD data $\langle q{\bar q}\rangle_f(T,eB)$ as
\ba
M_f(T,eB)&=&M_f(T=0,eB=0)\times \langle q{\bar q}\rangle_f(T,eB) + m_f(T,eB=0)
\nn\\
&\approx&M_f(T=0,eB=0)\times \langle q{\bar q}\rangle_f(T,eB)~.
\label{M_TB}
\ea
Using the LQCD data of $\langle q{\bar q}\rangle_f(T,eB)$ from Ref.~\cite{Bali1} and using those in Eq.~(\ref{M_TB}), we have plotted constituent quark mass $M_{f=u,d}$ against $T$-axis (left panel) and $eB$-axis (right panel) in Fig.~(\ref{fig:M_TB}). Here, reader can notice that constitute quark mass follow MC effect in low temperature domain and IMC effect near transition temperature domain. Enhancement of constituent quark mass by increasing values of $eB$ in low $T$-axis is noticed in left panel of Fig.~(\ref{fig:M_TB}) as well as in its right panel, we notice an increasing constituent mass curve with $eB$-axis at $T=100$ MeV. This enhancement of constituent quark mass with magnetic field in low temperature domain is proportionally mapping the MC effect of quark condensate. On the other hand, reduction of constituent quark mass by increasing values of $eB$ near transition temperature is noticed in left panel of Fig.~(\ref{fig:M_TB}) as well as in its right panel, we notice that constituent mass curve at $T=150$ MeV first increases then decreases with $eB$. This reduction of constituent quark mass with magnetic field near quark-hadron phase transition temperature domain is proportionally mapping the IMC effect of quark condensate. 

With this modified Lattice QCD inspired dispersion relation we have evaluated the Debye screening mass as % 
\begin{equation}\label{6}
m_{D}^{2}=g_s^2(T) T^2 \frac{N_c}{3} +\sum_f \dfrac{g_s^2(T)|q_feB|}{\pi^2T}
\int_{0}^{\infty} dp_z\sum_{l=0}^{\infty}(2-\delta_{l0})~{f}^l_q(\bar{E}_f^l)~\left(1-{f}^l_q(\bar{E}_f^l)\right),
\end{equation}
where we have considered 
%${f}_q^l(\bar{E}_f^l)$ is the modified quark distribution function and 
$g_s(T)$ - temperature dependent one loop running coupling~\cite{Haque:2018eph}:%{\color{red} (correct citation required from IN)}
%~\cite{Bannur:2006js,Zhu:2009zzi} {\color{red}change cite for one loop}.
%given as
\begin{eqnarray}
\label{as}
g_s^2(T)=4\pi \alpha_{s}(T)=\frac{24 \pi^2}{\left(11N_c-2 N_{f}\right)\ln \left(\frac{2\pi T}{\Lambda_{\overline{\rm MS}}}\right)}.
\end{eqnarray}

\subsection{Thermal Width and Binding Energy}
\label{sec:G_E}
Next, we focus on other relevant quantities like thermal width or dissociation rate, binding energy in the context of heavy quark potential. Their working formulas are described below.

Let us first discuss about thermal width or dissociation rate $\Gamma$, which can be formulated from the imaginary part of potential Im$~V(r,T,B)$, discussed in Sec.~(\ref{HQP}). In fact, the quantity Im $V(r,T,B)$ provide more detailed structure of dissociation rate along $r$, $T$ and $B$ axes. Now, if we know the $r$-profile of quarkonia wave function $\Psi({{r}})$, then the thermal width $\Gamma$ can be computed as~\cite{Agotiya:2016bqr}
\ba
\Gamma(T,B) = - \int d^3{\bf{r}}\, \left|\Psi({{r}})\right|^2{\rm{Im}}~V({\hat{r};T,B})~,
\label{Gamma}
\ea
where imaginary part of the potential is folded by the unperturbed (1S) Coulomb wave function, owing to the first-order perturbation theory.
%Decay width for the resonance state in thermal medium will reduce because of the effect of non-perturbative contributions coming from the string terms.
Here we take $\Psi(r)$ as the Coulombic wave function for ground state ($1s$,corresponding to $n=1$ ($J/\psi$ and $\Upsilon$))
given as
\ba
\Psi_{1s}(r)&=& \frac{1}{\sqrt{\pi r_B^3}}e^\frac{-r}{r_B}\nn
\label{psi}
\ea
where, $r_B$=$2/(\alpha_s ~m_Q)$ is the Bohr radius of the quarkonia system. By solving Eq.\ref{Gamma}, we have~\cite{Agotiya:2016bqr}
\begin{eqnarray}
\Gamma(T,B) =\left(1+\frac{3\sigma }{\alpha_s m_Q^2}\right)~\frac{4T}{\alpha_s }{\frac{m_D^2 }{m_Q^2}}~\log\frac{\alpha_s m_Q}{2m_D}~.
\label{Gm_TB}
\end{eqnarray}
Here also, one can see that the $T$ and $B$ dependent profile of $\Gamma$ is coming through $m_D(T,B)$. 
%%%%%%%%%%%%%%%%%%%%%%%%%%%%%%%%%%%%%%%%%%%%%%%%%%%%%%%%%%%%%%%%%%%%

%\subsection{Binding Energy}
%The potential is the function of only the radial coordinate 
%and we only have to solve the ordinary differential 
%equation of the radial part of the wave function. 
%
%The time-independent 
%Schr\"{o}dinger equation for the radial wave function reads 
%\begin{equation}
%- \frac{1}{2 m_q}
%\left( 
%\psi'' (r) + \frac{2}r \psi' (r)
%- \frac{ \ell (\ell+1)}{r^2} \psi (r)
%\right)
%+ 
% \Re V(r) \,  \psi  (r)
% = 
%\epsilon_{_{n\ell}} \,  \psi(r) ,
%\end{equation}
%where $m_q$ is the mass of the quarkonium system. 
% 
%We numerically solve this using the real part of the potential ~\ref{repotn}
%
Following Ref.~\cite{Agotiya:2008ie}, we can consider simple Culombic type
real part of heavy quark potential and then by solving its corresponding Schr\"{o}dinger equation, we can get the eigenvalues for charmonium and bottomonium system as
 \begin{eqnarray}
\label{bind1}
E_n=-\frac{1}{n^2} \frac{m_Q\sigma^2}{m^4_D}~,
\label{BE_TB}
\end{eqnarray}
where $m_Q$ is the mass of the heavy quark $c$ and $b$ respectively. Now its ground state eigen value for $n=1$ can be considered as binding energies of lowest possible charmonium and bottomonium states i.e. $\jpsi$ and $\Upsilon$ respectively. Similar to thermal width $\Gamma(T,B)$, binding energy $E_{n=1}(T,B)$ of $\jpsi$ and $\Upsilon$ will carry magnetic field dependent information through Debye mass $m_D(T,B)$. 
%%%%%%%%%%%%%%%%%%%%%%%%%%%%%%%%%%%%%%%%%%%%%%%%%%%%%%%%%%%%%%%%%%%

\section{Results}
\label{sec3}
In this section we will show and discuss about our results corresponding to heavy quark potential in a magnetized medium incorporating IMC based quark condensate. In the present work we have used $N_c=3, N_f=2$ and $\Lambda_{\overline{\rm MS}} = 0.176$ GeV~\cite{Haque:2014rua},
the string tension, $\sigma = 0.184 ~GeV^2$~\cite{nilima_aniso} and the value of $\alpha_{s}$ from Equation ~\ref{as}.

\begin{figure} %[tbh]
\begin{center}
\includegraphics[width=7.7cm]{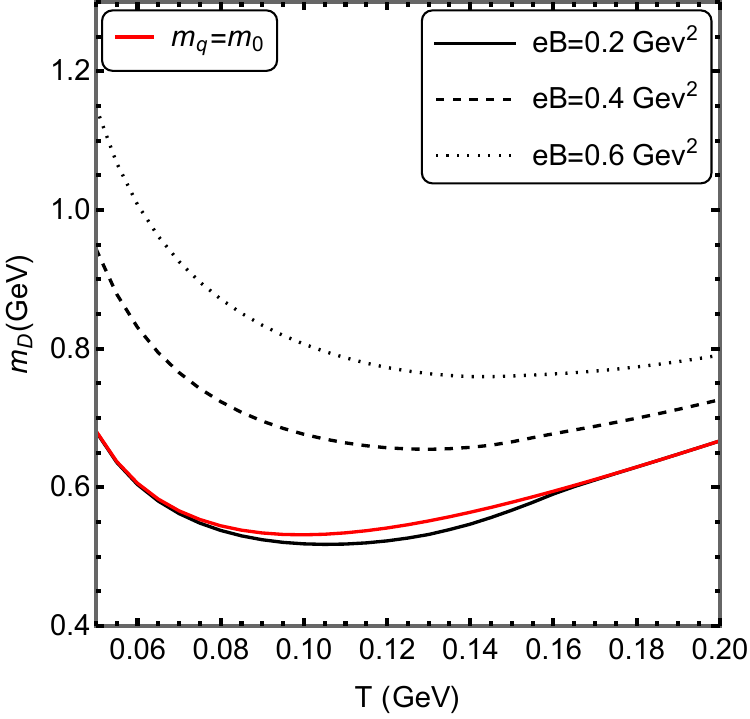} \includegraphics[width=7.7cm]{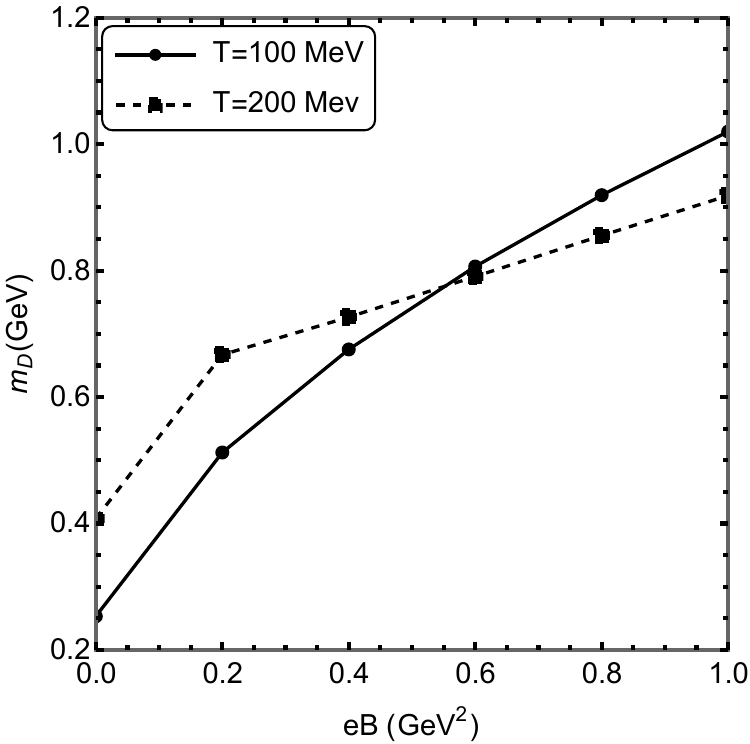}
\caption{Variation of the Debye mass ($ m_D $) with temperature for different values of magnetic fields $eB=0.2$ (black solid line), $0.4$ (dash line), $0.6$ (dotted line) GeV$^2$ in left panel and with magnetic field for different values of temperatures $T=100$ (black solid line), $200$ (dash line) MeV in right panel. Red solid line in left panel denotes for massless quark case.}
\label{mD}
\end{center}
\end{figure}

In the formalism part, we already discussed about the Fig.~\ref{fig:M_TB}, which represent graphically $T$ and $eB$ profile of constituent quark mass, based on LQCD quark condensate data. So, we will start from next step quantity - Debye mass, by using that $T$, $eB$ dependent constituent quark mass, which carry MC and IMC profiles.
In the left panel of Fig.~(\ref{mD}), we have plotted the variation of the Debye mass with $T$ for different values of magnetic field $eB=0.2$~GeV$^2$ (black solid line), $0.4$ (dash line), $0.6$ (dotted line) GeV$^2$. 
Whearas, the right panel of the Fig.~(\ref{mD}) shows the variation of Debye mass with $eB$ for different values of temperature $ T=100$ (black solid line), and $200$ (dash line) MeV.
If we follow the Debye screening mass expression, given in Eq.~(\ref{6}), then we can identify the $T$ and $eB$ dependent mathematical components. Sources of $T$ are FD distribution function $f^l_q(T)$ and coupling constant $g_s(T)$, which provide increasing and decreasing $T$ profile. Hence, due to their collective role, we notice that $m_D(T)$ decreases first and then increase. On the other hand, we can notice a straight forward $m_D\propto eB$ relation in Eq.~(\ref{6}). However, right panel of Fig.~(\ref{mD}) shows a deviation from the proportional relation, which is because a non-trivial $eB$ dependency is entering through the $eB$ dependent constituent quark mass $M(eB)$, located in FD distribution function. Also we notice that the effect of magnetic field is stronger at a lower temperature and becomes weaker at a higher temperature. 
%%%%%%%%%%%%%%%%%%%%%%%%%%%%%%%%%%%%%%%%%%%%%%%%%%%%%%%%%%%%%%%%%%%
%%%%%%%%%%%%%%%%%%%%%%%%%%%%%%%%%%%%%%%%%%%%%%%%%%%%%%%%%%%%%%%%%%%%

Now, if we compare with earlier $m_D(T,B)$ calculations, done by Refs.~\cite{Bandyopadhyay:2016fyd,Bonati:2017uvz,Singh:2017nfa,Kurian:2017yxj,Mitra:2017sjo,Kurain:2019,Karmakar:2019tdp,SG_VC}, then one can find that the new aspect of present work is the adoption of IMC based constituent quark mass. Though quark condensate and constituent quark mass are the good quantities, where we can see the IMC effect near transition temperature, but for the quantity like Debye mass $m_D(T,B)$, it is not zoomed in as it becomes fade due to integration of thermal distribution. For thermodynamical quantities like pressure, energy density or transport coefficients like electrical conductivity~\cite{Aritra_condB}, we get similar fact. So we will continue to present our final results as a IMC based outcome instead of zooming the actual modification, which is quite mild in graphical presentation. For seeing this modification in the left panel of Fig.~(\ref{mD}), we have plotted red solid line by considering massless quark case
%current quark mass $m_f=10$ MeV (close to massless quark limit) 
and compared it with black solid line, which has considered IMC based constituent quark mass $M_f(T,B)$. Within the $T=0.080$-$0.160$ GeV, a mild suppression of Debye mass is noticed due to considering the IMC based constituent quark mass $M_f(T,B)$ in place of current quark mass. This suppression of Debye mass is connected with the magnetized quark condensate in non-perturbative QCD (non-pQCD) domain, obtained from LQCD calculation~\cite{Bali1,Bali2}, which reveals IMC effect near transition temperature. So, this effect will propagate to other quantities of heavy quark phenomenology, as discussed next.      

%%%%%%%%%%%%%%%%%%%%%%%%%%%%%%%%%%%%%%%%%%%%%%%%%%%%%%%%%%%%%%%%%%%
\begin{figure} %[tbh]
\begin{center}
\includegraphics[width=7.7cm]{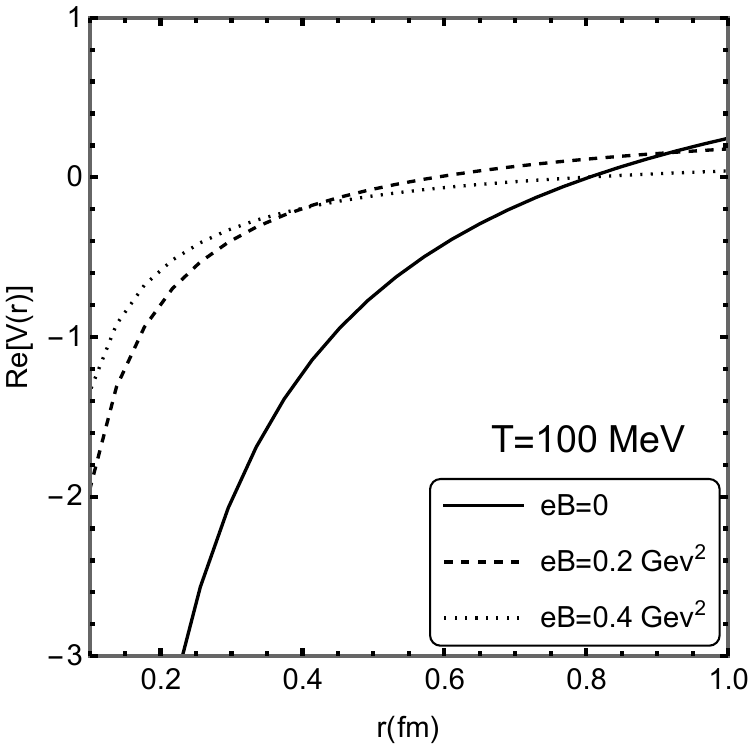}
\includegraphics[width=7.7cm]{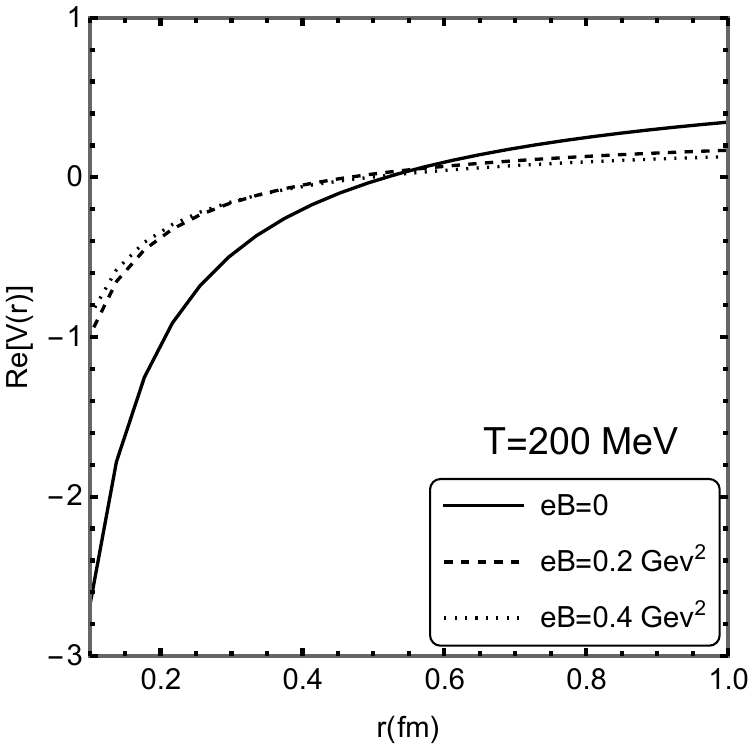}
\caption{Variation of the real part of potential with separation distance $ r $ between $Q\bar{Q}$ for three different values of the magnetic field and with fixed temperature $T=100$ MeV (left) and $T=200$ MeV (right).}
\label{repot}
\end{center}
\end{figure}
%%%%%%%%%%%%%%%%%%%%%%%%%%%%%%%%%%%%%%%%%%%%%%%%%%%%%%%%%%%%%%%%%%%%
We had plotted the variation of the real part of the potential  with the separation distance ($ r $) between the Q$\bar{Q}$ pair for different values of magnetic field (eB= $0.2 ~GeV^2$ and $0.4 ~GeV^2$) at $ T= 100 $ MeV (left) and $ T= 200 $ MeV (right) in Figure~\ref{repot} and compared it with the  absence of magnetic field i.e $eB= 0$. 
 From the figure~\ref{repot} we find that the screening is increasing with the increase in magnetic field and temperature. One can notice that the exponential decay with distance become more at a higher temperature ($ T= 200 $ MeV in right panel) as compared to lower temperature ($ T= 100 $ MeV in left panel). By shifting zero to non-zero magnetic field and increasing its values, similar dominancy in exponential behavior is noticed. It is because the exponential decay term is controlled by Debye mass, which increases with $T$ as well as $eB$. So increasing of screening with $T$ and $eB$ indicates loosely bound of quarkonium state at high $T$ and $eB$. 
%as compared to the lower temperatures which consequently results in easily dissociation. On the other hand, we can say that as the temperature increases subsequently the gluonic contribution becomes more which finally results in more screening.
%
%We also observed that the real part of potential without the magnetic field decreases slightly as compared to the one in presence of magnetic field.
%%%%%%%%%%%%%%%%%%%%%%%%%%%%%%%%%%%%%%%%%%%%%%%%%%%%%%%%%%%%%%%%%%%
\begin{figure}  %[tbh]
\begin{center}
\includegraphics[width=7.7cm]{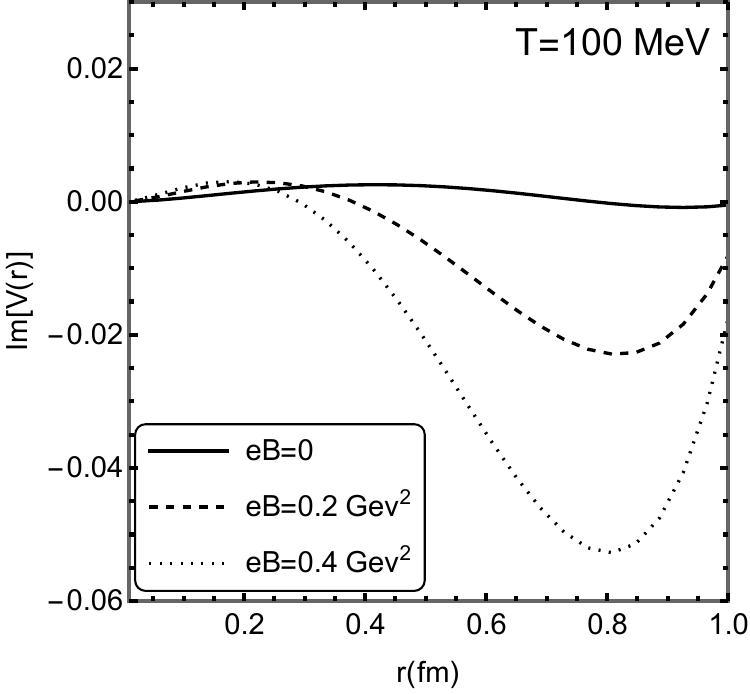}
\includegraphics[width=7.7cm]{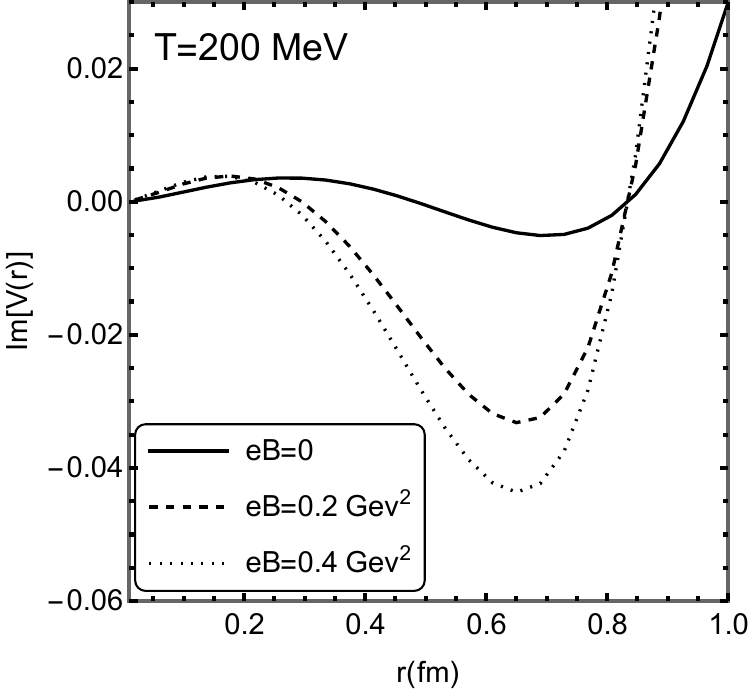}
\caption{Variation of the imaginary part of potential with separation distance $ r $ between $Q\bar{Q}$ for various values of magnetic field $T=100$ MeV (left) and $T=200$ MeV (right).}
\label{impot}
\end{center}
\end{figure}
%%%%%%%%%%%%%%%%%%%%%%%%%%%%%%%%%%%%%%%%%%%%%%%%%%%%%%%%%5%%%%%%%%

Similar to the real-part of the potential we have plotted the imaginary part of the potential with the separation distance ($ r $) for different values of magnetic field (eB= $0.2 ~GeV^2$ and $0.4 ~GeV^2$) at $ T= 100 $ MeV (left) and $ T= 200 $ MeV (right) in Fig.~\ref{impot} and compared it with the  absence of magnetic field i.e $eB= 0$. As we can see from the figure that magnitude of the imaginary part of the potential increases with the increase in magnetic field and hence it provides more contribution to the thermal width obtained from the imaginary part of the potential. We also find that the magnitude of imaginary part of potential is more at a higher temperature ($T=200$ MeV) as compared to a lower temperature ($T=100$ MeV) for a given $eB$ and $r$. This observation is again coming from the fact that Debye mass which is a function of temperature and magnetic field is found to be increased with temperature and magnetic field. So, grossly we notice that real and imaginary part of heavy quark potential are modified with $T$ and $eB$ due to non-trivial profile of LQCD based condensate and their modifications approach towards more screening and dissociation with increasing of $T$ and $B$. It indicates a possibility of low dissociation temperature due to magnetic field, which we will explicitely see latter.
%
%We also observed that the imaginary part of potential without the magnetic field decreases slightly as compared to the one in presence of magnetic field.

Next, let us integrate the r dependence of imaginary and real potential through Coulomb-type probability distribution function and
proceed to compute thermal width or dissociation probability $\Gamma(T,B)$ by using Eq.~(\ref{Gm_TB}) and binding energy using the Eq.~(\ref{BE_TB}). Imaginary and real part of heavy quark potential are their respective sources, whose $T$ and $eB$ profiles are mainly coming from $m_D(T,eB)$.
%%%%%%%%%%%%%%%%%%%%%%%%%%%%%%%%%%%%%%%%%%%%%%%%%%%%%%%%%%%%%%%%%%%%
\begin{figure}  %[tbh]
\begin{center}
\includegraphics[width=7.6cm]{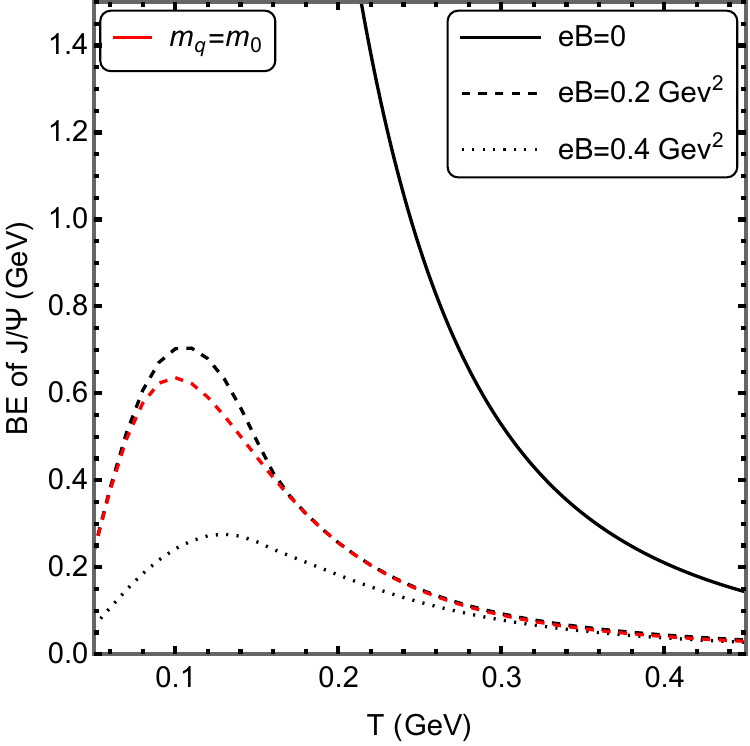}
\includegraphics[width=7.4cm]{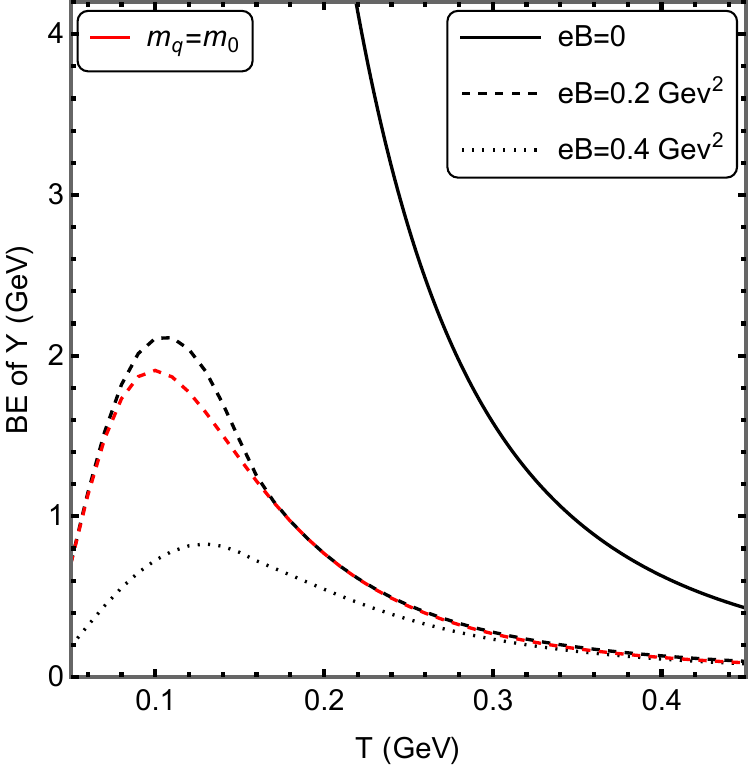}
\caption{Variation of binding energy(in $GeV$) with temperature with different values of magnetic field (eB) for $J/\psi$ (left panel) and $\Upsilon$ (right panel).}
\label{be}
\end{center}
\end{figure}
Let us first discuss about binding energy (BE), plotted in Fig.\ref{be}. Its left panel shows the binding energies of $J/\psi$ as a function of $T$ for various values of magnetic field (eB=$0.2 ~GeV^2$ and $0.4 ~GeV^2$) and compared it with the results, considering current quark mass (red line). We observe that binding energy is decreasing as the temperature and magnetic field both are increasing.
In the right panel of Fig.\ref{be} we have plotted the  binding energies of $\Upsilon$ as a function of $T$ for various values of magnetic field (eB=$0.2 ~GeV^2$ and $0.4 ~GeV^2$) and compared it with without magnetic field ($eB=0$).The similar behaviour has been observed for $\Upsilon$ also, except that the value of binding energy for $J/\Psi$ is lower as compared to the value for $\Upsilon$, which is due to their mass difference. We have taken 
(charmonium) $J/\psi$ and (bottomonium) $\Upsilon$ masses as $m_c=1.5$ GeV and $m_b=4.5$ GeV respectively.
%%%%%%%%%%%%%%%%%%%%%%%%%%%%%%%%%%%%%%%%%%%%%%%%%%%%%%%%%%%%%%%%%%%%
\begin{figure} %[tbh]
\begin{center}
\includegraphics[width=5.4cm]{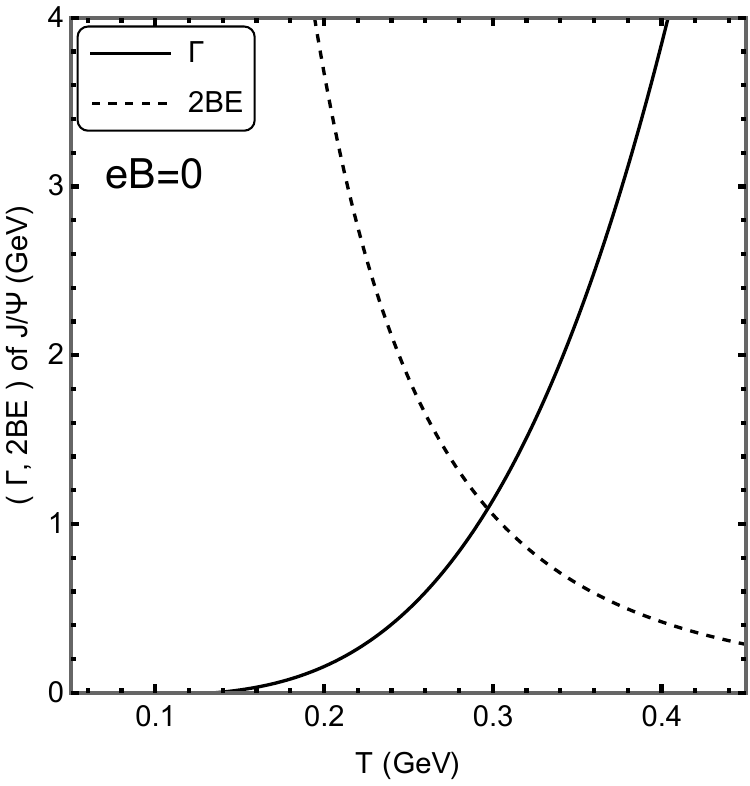}
\includegraphics[width=5.3cm]{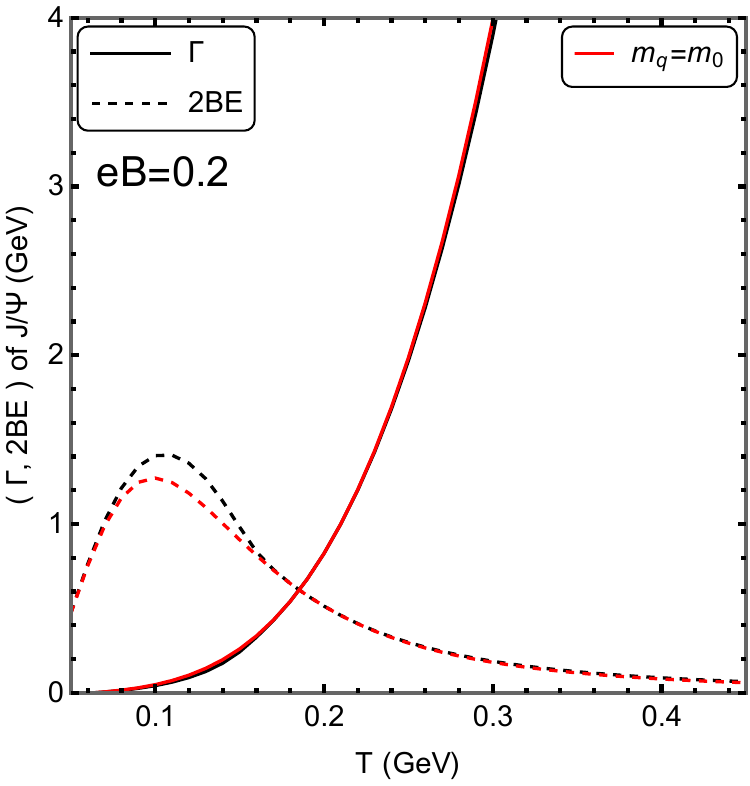}
\includegraphics[width=5.3cm]{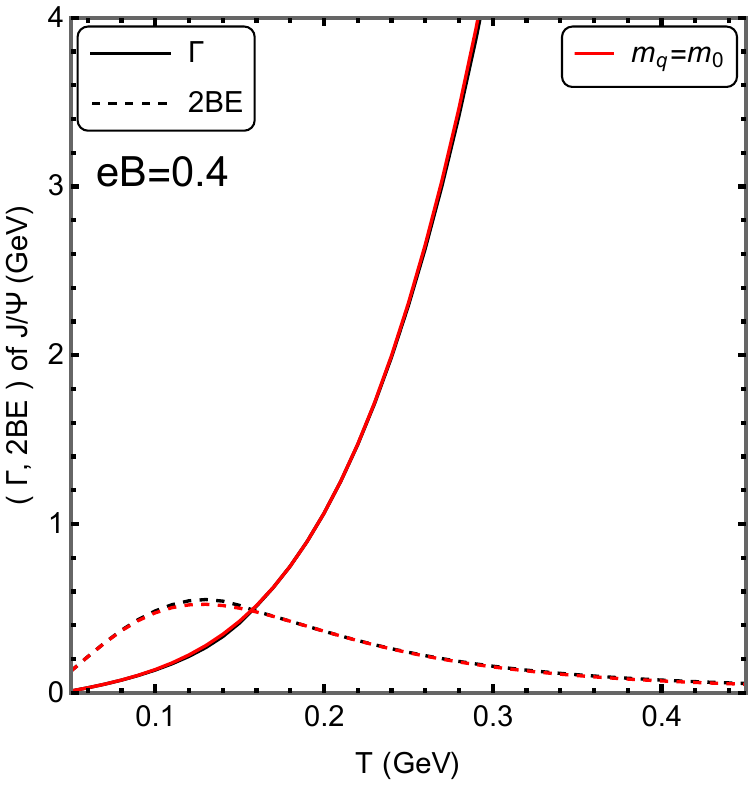}
\caption{ Variation of $\Gamma$, 2BE($E_b$) with Temperature with different values of magnetic field (eB) for $J/\psi$}
\label{gcharm}
\end{center}
\end{figure}
%%%%%%%%%%%%%%%%%%%%%%%%%%%%%%%%%%%%%%%%%%%%%%%%%%%%%%%%%%%%%%%%%%%
%%%%%%%%%%%%%%%%%%%%%%%%%%%%%%%%%%%%%%%%%%%%%%%%%%%%%%%%%%%%%%%%%%%%
\begin{figure}  %[tbh]
\begin{center}
\includegraphics[width=5.4cm]{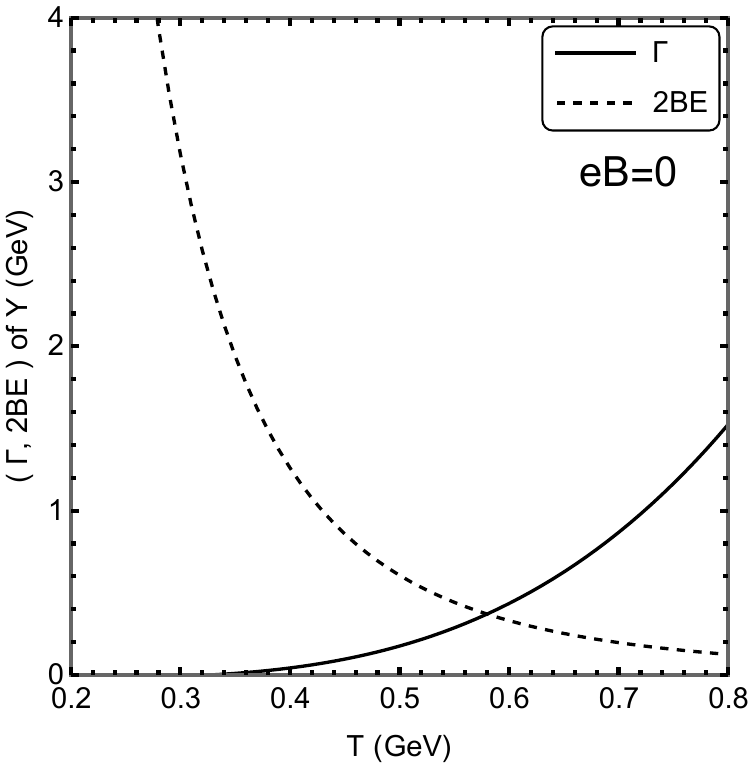}
\includegraphics[width=5.3cm]{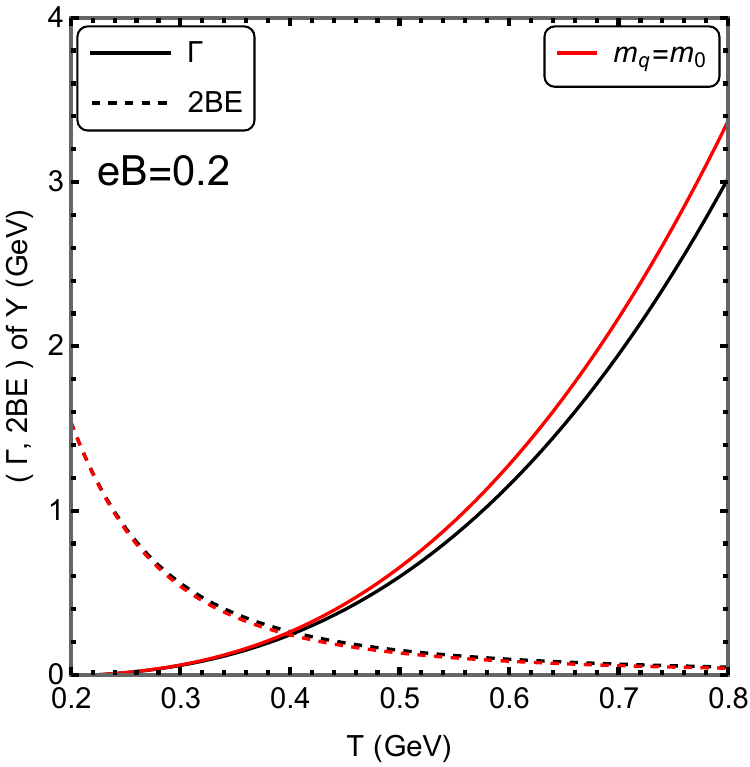}
\includegraphics[width=5.3cm]{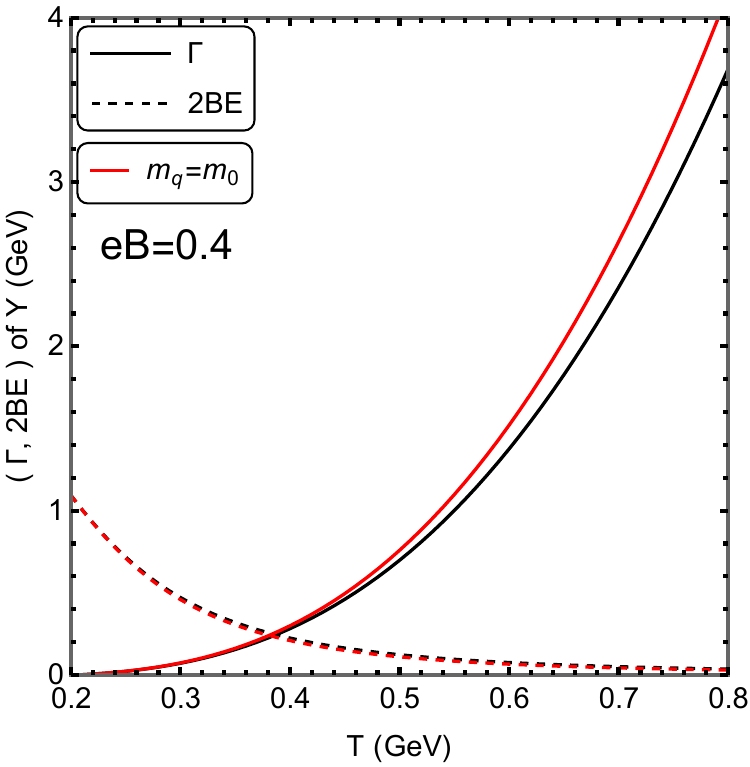}
\caption{Variation of $\Gamma$, 2BE($E_b$) with Temperature with different values of magnetic field (eB) for $\Upsilon$}
\label{gupsilon}
\end{center}
\end{figure}
%%%%%%%%%%%%%%%%%%%%%%%%%%%%%%%%%%%%%%%%%%%%%%%%%%%%%%%%%%%%%%%%%%%
 %We also found that binding energy in absence of magnetic field is greater than the binding energy in presence of magnetic field. This is because in the presence of magnetic field, real-part of the potential becomes more screened due to the presence of temperature and magnetic dependent debye mass term as compared to one in absence of magnetic field. This observation is similar for both $J/\psi$ and $\Upsilon$ cases.

Next, let us come to the estimation of thermal width or dissociation probability, given in Eq.~(\ref{Gm_TB}), which is obtained by substituting Eq.~(\eqref{imis}) in Eq.~(\eqref{Gamma}). So Eq.~(\eqref{imis}) represents a detailed tomography of dissociations of quarkonia state, while Eq.~(\eqref{Gamma}) present its integrated values.  
%As given in the reference~\cite{utt} the quarkonia state will dissolve whenever the thermal width $\Gamma$ of the given quarkonium is as large as  twice the real part of the binding energy.
%
Now this heavy quark bound states, called quarkonia, can be dissociate completely, when thermal width will be greater than twice of the the binding energy i.e. $\Gamma\geq 2BE$~\cite{utt}. Now, if one plots twice the binding energy ($2BE$) along with the thermal width ($\Gamma$) against $T$-axis, then their intersection point can be considered as dissociation temperature ($T_d$).
%
%We have plotted the twice of binding energy along with the thermal
%width and obtain the dissociation temperature as a point
%of their intersection
We have done this in fig.~\ref{gcharm} and in fig.~\ref{gupsilon} for ground state of charmonium and bottomonium respectively with different values of magnetic field.

If we analyse the  fig.~\ref{gcharm} and fig.~\ref{gupsilon} we can find that the thermal width is increasing with the increase in magnetic field. Also we can observe that the width for the  $J/\psi$ is much larger than  the $\Upsilon$, because charmonium states are larger in size and smaller in masses as compared to bottomonium states which are smaller in size and larger in masses and hence will get dissociated at higher temperatures. 
Interestingly, we see that $ \Gamma $ increases and BE decreases with magnetic field, which results in the early dissociation of quarkonium states.
Our findings of the dissociation temperatures from intersection points in graphs are enlisted in table ~\ref{tab:t2}. 
%in the absence and
%presence of weak magnetic field for $J/\psi$ and $\Upsilon$ after equating the thermal width with the twice of the binding energy. 
%The $J/\psi$ state is seen to be dissociated at $297 MeV$ for $eB=0$, $T=186 MeV$ for $eB=0.2$ and $T= 158 MeV$ for $eB=0.4$.
%Whearas $\Upsilon$ state is seen to be dissociated at $580 MeV$ for $eB=0$, $T=403 MeV$ for $eB=0.2$ and $T= 388 MeV$ for $eB=0.4$.
%
%We can observed that the dissociation temperatures become slightly higher in the absence of magnetic field as compared to the presence of magnetic field. As discussed earlier it is evident from the table that $\Upsilon$ is dissociating at higher temperature as compared to $J/\psi$.
%%%%%%%%%%%%%%%%%%%%%%%%%%%%%%%%%%%%%%%%%%%%%%%%%%%%%%%%%%%%%%%%%%%%
\begin{table}[tbh]
\begin{center}
\begin{tabular}{ |p{2cm}||p{2cm}|p{2cm}|p{2cm}|  }
\hline
State &$eB=0$ & $eB=0.2$&$eB=0.4$\\
\hline\hline
$\jpsi$&297 & 186 & 158 \\
\hline
$\Upsilon$&580 & 403 & 388 \\
\hline
\end{tabular}
\caption{The dissociation temperature($T_D$) for the
quarkonia states (in units of MeV), when $\Gamma$= 2BE} 
\label{tab:t2}
\end{center}
\end{table} 
%%%%%%%%%%%%%%%%%%%%%%%%%%%%%%%%%%%%%%%%%%%%%%%%%%%%%%%%%%%%%%%%%%%%
We notice that dissociation temperature of quarkonia states decreases as magnetic field increases. This fact is exactly similar with the reduction of quark-hadron phase transition temperature due to magnetic field, which is connected with IMC aspect of QCD. If QCD follow MC near transition temperature, then magnetic field will push the location of transition temperature towards the higher values. However, reduction of dissociation temperature with magnetic field can not be linked with IMC as constituent quark mass has very mild impact on Debye mass and heavy quark phenomenology. In other word, the fact of reduction of dissociation temperature with magnetic field remain same for both MC and IMC. This conclusion can be established from Figs.~(\ref{gcharm}) and (\ref{gupsilon}), where we notice that $T_d$ remain almost same for IMC-based constituent quark mass (black line) and massless quark mass case (red line).

\section{Conclusions}
\label{sec4}
%In the present theoretical study, we have explored the effects of strong magnetic field on the dissociation of quarkonia in a thermal QCD.
%We have taken only the LLL contribution in our calculation. 
%We have calculated debye mass with HLL corrections in the strong magnetic field. This debye screening mass can be directly evaluated from the temporal component of the gluon self energy tensor ($\Pi_{00}(p_0,\vec{p})$) by employing the static limit ($|\vec{p}|=0, p_0\rightarrow 0$) through a perturbative order by order evaluation.

We have revisited the medium modified heavy quark potential at finite magnetic field. This is done by obtaining the real and imaginary parts of the resummed gluon propagator, which in
turn gives the real and imaginary parts of the dielectric permittivity. A temperature and magnetic field dependent Debye mass, capturing the information of inverse magnetic catalysis, is entering into the gluon propagator.
Now the real and imaginary parts of the dielectric permittivity  will be used to evaluate the real and imaginary parts of the complex heavy quark potential. We notice that the Debye screening mass increases with temperature and magnetic field, so Debye exponential part of real part of potential become more dominant, which interpret more screening and favoring the dissociation process. With respect to earlier works, present work has incorporated two new ingredients - inverse magnetic catalysis information and all Landau level summations.
%
%We have studied the effects of strong magnetic field on the real and imaginary parts of the potential. We have found that the screening mass will be increasing with the increase in magnetic field i.e. screening will be more at a higher temperature $(T = 200 MeV)$ as compared to lower temperature $(T = 100 MeV)$. We also observed that the real part of potential without the magnetic field decreases slightly as compared to the one in presence of magnetic field. On the other hand, the magnitude of imaginary part of the potential with the separation distance (r) gets increased for increasing values of magnetic field. We also find that for a given r, imaginary part of potential is more at a higher temperature $(T = 200 MeV)$ as compared to a lower temperature $(T = 100 MeV)$. We also observed that the imaginary part of potential without the magnetic field decreases slightly as compared to the one in presence of magnetic field.
%

The real part of the potential is used in the time-independent Schr\"odinger equation for the radial wave function to obtain the binding energy of heavy quarkonia, whereas the imaginary part is
used to calculate the thermal width. We observe that the binding energies of $J/\psi$ and  $\Upsilon$ are decreasing and their thermal widths are increasing as the temperature and magnetic field both are increasing. These decreasing of binding energy and increasing of thermal width of heavy quarkonia will push its dissociation probability. By plotting the twice of binding energy along with the thermal width, one can obtain the dissociation temperature as a point of their intersection. We noticed that dissociation temperature of heavy quarkonia can be reduced due to magnetic field. 
%We also found that binding energy in absence of magnetic field is greater than the binding energy in presence of magnetic field. This is because in the presence of magnetic field, real-part of the potential
%becomes more screened as compared to one in absence of magnetic field.
%
%Finally we have plotted the twice of binding energy along with the thermal width and obtain the dissociation temperature as a point of their intersection. We find that the thermal width is increasing with the increase in magnetic field. Also we can observe that the width for the $J/\psi$ is much larger than the $\Upsilon$. We can observed that the dissociation temperatures become slightly higher in the absence of magnetic field as compared to the presence of magnetic field 
%for example, $J/\psi$ state is seen to be dissociated at $297 MeV$ for $eB=0$, $T=186 MeV$ for $eB=0.2$ and $T= 158 MeV$ for $eB=0.4$.
%Whearas $\Upsilon$ state is seen to be dissociated at $580 MeV$ for $eB=0$, $T=403 MeV$ for $eB=0.2$ and $T= 388 MeV$ for $eB=0.4$.
%
%Clearly, the present investigation is limited to very high magnetic field where only lowest Landau level contributes to the dielectric function. To extend the present work, we aim to incorporate
%the arbitrary magnetic field  and study the dissociation of heavy
%quarkonia. Another, interesting direction would be to couple the analysis to the physics of momentum anisotropy and instabilities in the early stages of the heavy-ion collisions in the presence of strong magnetic field and study its impact on the heavy quarkonia dissociation. 
 
\section*{Acknowledgement}
 I.N. acknowledges IIT Bhilai for the academic visit and hospitality during the course of this work. 
 A.B. acknowledges the support from Guangdong Major Project of Basic and Applied Basic Research No. 2020B0301030008 and Science and Technology Program of Guangzhou Project No. 2019050001. Authors are thankful to Mujeeb Hassan for useful discussion. I.N., A.B., R.G thank to Shivani Valecha, Sunima Baral, Srikanta Debata, Purushottam Sahu, Naba Kumar Rana for helping them in arranging accommodation.

\end{document}